\documentclass[a4paper,twocolumn,english,aps,accepted=2023-02-07]{quantumarticle}
\pdfoutput=1

\usepackage[utf8]{inputenc}
\usepackage[T1]{fontenc}
\usepackage{amsmath,amssymb,bbm,mathrsfs,bm,braket,mathtools,color,graphicx,comment}
\usepackage{xcolor}
\usepackage{bbold}
\usepackage{appendix}
\usepackage{hyperref}
\usepackage[mathscr]{euscript}
\usepackage{subfigure}
\usepackage{enumitem}
\usepackage{multirow}
\usepackage{cite}
\usepackage[labelfont=bf]{caption}
\begin{document}

\title{Parity Quantum Optimization:~Benchmarks}

\author{Michael Fellner}
\affiliation{Parity Quantum Computing GmbH, A-6020 Innsbruck, Austria}
\affiliation{Institute for Theoretical Physics, University of Innsbruck, A-6020 Innsbruck, Austria}

\author{Kilian Ender}
\affiliation{Parity Quantum Computing GmbH, A-6020 Innsbruck, Austria}
\affiliation{Institute for Theoretical Physics, University of Innsbruck, A-6020 Innsbruck, Austria}

\author{Roeland ter Hoeven}
\affiliation{Parity Quantum Computing GmbH, A-6020 Innsbruck, Austria}
\affiliation{Institute for Theoretical Physics, University of Innsbruck, A-6020 Innsbruck, Austria}

\author{Wolfgang Lechner}
\email{wolfgang@parityqc.com\\ wolfgang.lechner@uibk.ac.at}
\affiliation{Parity Quantum Computing GmbH, A-6020 Innsbruck, Austria}
\affiliation{Institute for Theoretical Physics, University of Innsbruck, A-6020 Innsbruck, Austria}

\begin{abstract}
We present benchmarks of the parity transformation for the Quantum Approximate Optimization Algorithm (QAOA). We analyse the gate resources required to implement a single QAOA cycle for toy models of real-world scenarios. In particular, we consider random spin models with higher order terms, as well as the problems of predicting financial crashes and finding the ground states of electronic structure Hamiltonians. For the spin models studied our findings imply a significant advantage of the parity mapping compared to the standard gate model in terms of the number of CNOT gates required on a square lattice with nearest-neighbor connectivity, at the cost of an increased number of qubits. Our results suggest that in combination with its full parallelizability of gates, the parity architecture has the potential to boost the race for demonstrating quantum advantage.
 
\end{abstract}
\maketitle

\section{Introduction}
The development of algorithms to solve hard optimization problems are a cornerstone of computer science. The repertoire of optimization techniques has recently been enhanced with the emergence of quantum algorithms and the recent progress in quantum hardware. A promising approach is the paradigm of Quantum Annealing \cite{kadowaki1998quantum,Albash2018}, an analog quantum computing method based on the quantum adiabatic theorem \cite{Born1928}. Quantum Annealing inspired the Quantum Approximate Optimization Algorithm (QAOA) \cite{Farhi2014}, which is considered a promising approach to show quantum advantage on near-term devices \cite{preskill2018quantum}. As a digital algorithm, it is implemented on gate-based quantum devices. QAOA has been the subject of many investigations in the last few years \cite{Hadfield2019, Brandao2018, Zhou2020, Willsch2020, Farhi2020}. One of the challenging aspects when implementing these quantum algorithms on physical quantum devices is the non-locality of interactions on the chip. For example, many NP-hard optimization problems map to an Ising Hamiltonian with long-range connectivity, constraints and higher-order interactions which cannot be directly embedded on a quantum chip \cite{Lucas2014}. Thus, benchmarks of encoding of optimization problems on realistic hardware are a pressing need for the development of next generation quantum devices. 

In this paper, we present benchmarks comparing the parity architecture to standard gate model encodings on the same hardware for toy models of real-world use cases as well as generic problems. We compare the number of gates required in three scenarios: a standard compiler using CNOT gates on a square lattice, the parity compiler on the same hardware with the same gates, and the parity compiler using parity gates (i.e., 4-body gates). The parity architecture~\cite{compilerpaper}, which generalizes the LHZ-architecture \cite{Lechner2015}, allows one to implement problems with highly non-local higher-order terms, using only local interactions, and to encode constraints \cite{constraintpaper}. The mapping is thus an alternative to expressing problems as quadratic unconstrained binary optimization (QUBO) problems and allows for a direct implementation of higher-order constrained optimization (HCBO) problems. The LHZ-architecture, and to a large extent also the generalized parity scheme, can be implemented with fully parallelizable gates  \cite{Lechner2018}  which will open completely new directions in the design of quantum gates (e.g.\ Ref.~\cite{rydbergpaper}).  

We will first discuss general aspects about the number of CNOT gates required to encode problems with higher-order terms on a two-dimensional chip using standard compilers. Then, from a large variety of possible applications in different disciplines \cite{Perdomo-Ortiz2012, Stollenwerk2020, Xia2018, Orus2019, Jiang2018}, we choose to investigate two problems. First, we focus on the prediction of a financial crash, which is an important issue in finance and economics. Second, we present a benchmark for calculating the ground state of electronic structure problems, representing a prominent problem in quantum chemistry that is infeasible with classical resources.

In the following section, we elaborate on the underlying concepts of the benchmarks, which include a more detailed description of QAOA and the parity compiler. After that, we introduce the optimization problems to investigate and describe the corresponding benchmarks. Finally, we present the implications and conclusions we can draw from our benchmarks.
\section{Background}

\subsection{Parity Quantum Computing}
The parity transformation, described in Ref.~\cite{compilerpaper} and Ref.~\cite{constraintpaper} of this series is the generalization of the LHZ-architecture \cite{Lechner2015}. For completeness we summarize the main steps. Consider the general problem Hamiltonian for $N$ qubits
\begin{equation}\label{eq:logical_hamiltonian}
    \begin{aligned}
            H_p&=\sum_{i=1}^N J_i \sigma_z^{(i)}+ \sum_{i=1}^N\sum_{j>i} J_{ij} \sigma_z^{(i)}\sigma_z^{(j)}+ \\
            &+ \sum_{i=1}^N\sum_{j>i}\sum_{k>j} J_{ijk} \sigma_z^{(i)}\sigma_z^{(j)}\sigma_z^{(k)}+ \dots, 
    \end{aligned}
\end{equation}
containing spin terms of arbitrary order. Let us denote the number of terms in $H_p$ by $K$. 
The parity compiler relies on the mapping of such a general interaction Hamiltonian for $N$ qubits
to a Hamiltonian involving \textit{physical qubits} $\Tilde{\sigma}$ which represent all the products of the logical qubit operators involved in Eq.~\eqref{eq:logical_hamiltonian}. As usually ${K>N}$, the Hilbert space of the new Hamiltonian is larger than the Hilbert space of the logical problem. Therefore, $N_C={K-N+D}$ constraints have to be imposed in order to preserve the number of degrees of freedom, where $D$ denotes the number of degeneracies in the original Hamiltonian. These constraints are chosen such that invalid configurations of the physical qubits are energetically unfavorable, using generalized closed cycles in the logical hypergraph \cite{compilerpaper}. With that, the physical Hamiltonian reads
\begin{equation}\label{eq:physical_hamiltonian}
    H_\text{phys}=\sum_{k=1}^K J_k \Tilde{\sigma}_z^{(k)}+\sum_{l=1}^{N_C}c_l \Tilde{\sigma}_z^{(l, 1)}\Tilde{\sigma}_z^{(l, 2)}\Tilde{\sigma}_z^{(l, 3)}\left[\Tilde{\sigma}_z^{(l, 4)}\right],
\end{equation}
where the first term sums over all physical qubits representing an interaction, and the second term enforces the constraints. The coefficients $c_l$ denote the constraint strength and the brackets around $\Tilde \sigma_z^{(l, 4)}$ indicate that a constraint can contain either 3 or 4 qubits and will be omitted in the remainder of this paper. A constraint can be any product of $k$-body terms. However, for two-dimensional lattices, 3- and 4-body terms are the most practical, because it is crucial for implementations that constraints are \textit{geometrically local} interactions, e.g.\ nearest neighbors on a square lattice. 

For Hamiltonian quantum computing applications, such as adiabatic quantum computing, the constraints are implemented on so-called plaquettes. A plaquette is a set of 3 or 4 qubits arranged on the vertices of a square, such that an interaction between them can be implemented. For a more detailed description see Ref.~\cite{compilerpaper}. In digital devices, the gate model allows for more flexibility and the qubits involved in a constraint are placed contiguously on the hardware grid such that the corresponding QAOA unitary can be implemented with consecutive CNOT gates. The plaquettes can be viewed as a special case of such lines. 

\subsection{Quantum Approximate Optimization}
The Quantum Approximate Optimization Algorithm (QAOA) \cite{Farhi2014} is a variational hybrid quantum-classical approach to tackle combinatorial optimization problems and is considered one of the most promising candidates to demonstrate quantum advantage in near-term devices \cite{preskill2018quantum,bharti2021noisy}. In order to find approximations to the ground state of a system, we prepare a candidate state by applying parametrized unitary operators to some initial state, using a quantum computer. In a classical feedback loop, the energy expectation value of the final state is evaluated and parameters are updated. This is repeated until a stopping criterion is reached.

In more detail, suppose we are interested in the ground state of a problem Hamiltonian $H_{\rm{p}}$. We define two unitary operators,
\begin{equation}\label{eq:driver_unitary}
    U_x(\beta) = \prod_{j=1}^N e^{-i\beta \sigma_x^{(j)}},
\end{equation}
which is referred to as the \textit{driver unitary} and 
\begin{equation}\label{eq:problem_unitary}
    U_{\rm{p}}(\gamma) = e^{-i\gamma H_{\rm{p}}},
\end{equation}
which is called the \textit{phase-separation unitary}. Here, ${\beta\in [0, \pi)}$ and ${\gamma\in [0, 2\pi)}$ \cite{Farhi2014}. In the spirit of QAOA, a solution candidate $\ket{\psi}$ for the ground state of $H_{\rm{p}}$ is prepared by sequentially applying the operators \eqref{eq:driver_unitary} and \eqref{eq:problem_unitary} $p$ times to some initial state $\ket{\Psi_0}$, i.e.\
\begin{equation}
    \ket{\psi}=U_x(\beta_p)U_{\rm{p}}(\gamma_p)\dots U_x(\beta_1)U_{\rm{p}}(\gamma_1)\ket{\Psi_0}.
\end{equation}
The state $\ket{\psi}$ is prepared sufficiently many times to obtain a reasonable estimate for the expectation value $\braket{\psi|H_{\rm{p}}|\psi}$. The expectation value is passed to a classical optimizer which returns new parameters $\beta_i$ and $\gamma_i$ (${1\leq i \leq p}$) and the procedure is repeated.

The integer $p\geq 1$ is referred to as the \textit{QAOA depth} or the \textit{number of QAOA cycles}. The initial state is usually chosen to be the equal superposition of all computational basis states, i.e.\ 
\begin{equation}\label{eq:initial_state}
    \ket{\Psi_0}=\frac{1}{\sqrt{2^N}}\sum_{z\in\{0, 1\}^N}\ket{z}=\bigotimes_{i=1}^N \ket{+}_i.
\end{equation}
It has been proven that for $p\to\infty$ the procedure will converge to the correct ground state \cite{Farhi2014}. For some problem types, such as the Max-Cut problem, there are performance guarantees even for low QAOA depths \cite{Farhi2014, Wurtz2021}. Recent research also indicates \textit{parameter concentration} in the QAOA protocol which means that good QAOA parameters found for (small) problem instances are also adequate for other, potentially larger instances \cite{Zhou2020, Akshay2021, Brandao2018, Streif2020}. This would significantly reduce classical as well as quantum resources needed for the parameter optimization.

For QAOA in the parity scheme \cite{Lechner2018}, the problem Hamiltonian \eqref{eq:physical_hamiltonian} can be split into the local field part ${H_z=\sum_{k=1}^K J_k \Tilde{\sigma}_z^{(k)}}$ and the constraint part ${H_C=\sum_l^{N_C} c_l \Tilde{\sigma}_z^{(l, 1)}\Tilde{\sigma}_z^{(l, 2)}\Tilde{\sigma}_z^{(l, 3)}\Tilde{\sigma}_z^{(l, 4)}}$. Here, the operators $\Tilde{\sigma}_z^{(l, i)}$ denote the Pauli $z$-operators acting on the $i$-th qubit in constraint $l$.
Therefore, we can split the problem unitary by introducing another set of $p$ QAOA parameters ${{\Omega_1, \dots, \Omega_p}}$, such that we obtain
\begin{equation}
    U_z(\gamma)= \prod_{k=1}^K e^{-i\gamma J_k \Tilde{\sigma}_z^{(k)}}
\end{equation}
and
\begin{equation}
    U_c(\Omega)=\prod_{l=1}^{N_C} e^{-i \Omega c_l \Tilde{\sigma}_z^{(l, 1)}\Tilde{\sigma}_z^{(l, 2)}\Tilde{\sigma}_z^{(l, 3)}\Tilde{\sigma}_z^{(l, 4)}},
\end{equation}
where $\Omega\in [0, 2\pi)$. An even more advanced approach could be to also optimize the coefficients $c_l$ as QAOA parameters \cite{Lechner2018}. For our purposes, however, we set ${c_l=-1}$ for all $l$ for the sake of simplicity. With these unitaries, the QAOA protocol reads
\begin{equation}
\begin{aligned}
\ket{\psi}=U_x(\beta_p)&U_c(\Omega_p)U_z(\gamma_p)\times \cdots\\ \cdots\times &U_x(\beta_1)U_c(\Omega_1)U_z(\gamma_1)\ket{\Psi_0}.    
\end{aligned}
\end{equation}
Note that the initial state and the driver unitary [see Eqs.~\eqref{eq:driver_unitary} and \eqref{eq:initial_state}] have to be defined on $K$ instead of $N$ qubits in this context.
This protocol enhances the flexibility of the algorithm at the cost of enlarging the search space for the classical parameters.

\section{Embedding Strategies}
Here, we outline different strategies to embed optimization problems on digital quantum devices for QAOA. This includes the implementation in the standard gate model as well as in the parity architecture.

\subsection{QAOA in the standard gate model}
In order to perform QAOA without using the parity mapping (PM), the (potentially high-order-) terms in the problem Hamiltonian have to be implemented as a QAOA problem unitary as described in the introductory section. This results in terms of the form
\begin{equation}
    U_P = \prod_\text{interactions}\exp\left(i\alpha \prod_k J_k \sigma_z^{(k)}\right),
\end{equation}
where the first product runs over all interaction terms in the problem Hamiltonian and the product in the exponent
runs over all qubits involved in the interaction.
In this paper, we consider square lattice chips with nearest-neighbor connectivity. Such qubit lattices are provided by state-of-the-art quantum devices \cite{Bloch2008, Saffman2010, Henriet2020, Arute2019}. The unitary operator corresponding to an $n$-body interaction is implemented with ${2(n-1)}$ CNOT gates and a single $R_z$ rotation. Such a circuit element is also referred to as a \textit{phase gadget} and can be implemented with several CNOT arrangements \cite{Cowtan2020, DeGriend2020, Sivarajah2020}. As there might be non-local interactions in the problem Hamiltonian (i.e.\ interactions not neighboring on the square lattice), additional SWAP gates are required in the direct implementation. Note that a single SWAP-gate consists of 3 CNOT gates, which makes it very resource-intensive.
With that, the number of CNOT gates required for the implementation depends on
\renewcommand{\labelenumi}{(\roman{enumi})}
\begin{enumerate}
    \item the number of terms in the problem Hamiltonian,
    \item their interaction order and
    \item the locality of the qubits involved in each interaction.
\end{enumerate}

The direct implementation of interactions becomes especially expensive whenever there are highly non-local interactions and
high-order interactions involved in the problem. The non-locality of interactions can be influenced, to some extent, by modifying the arrangement of qubits on the chip and by the connectivity of the chip.

\subsection{Parity QAOA with CNOT gates}
In the parity mapping, one possibility to perform QAOA is to impose the constraints by CNOT gates in the same manner as for the many-body terms in the standard gate model (GM). The number of CNOT gates required for one QAOA cycle is therefore, up to a factor, given by the number of parity constraints $N_C$ and does not depend on the interaction order of the terms or on their locality. This is because the interactions in the parity mapping are local by definition~\cite{Lechner2015}. If no ancilla qubits are used in the parity compilation (which we assume in the benchmarks presented here), the relation
\begin{equation}\label{eq:n_constraints}
N_C=K-N
\end{equation}
usually holds. Here, $K$ is the number of terms involved in the Hamiltonian and $N$ the number of logical qubits. More constraints can be necessary if additional symmetries occur in the logical system. For example, unless at least one $k$-body interaction with odd $k$ exists, there is a global spin-flip symmetry in the logical problem. The implementation procedure of the constraints in the parity scheme is the same as for higher-order interactions in the original Hamiltonian, as constraints are just 3- or 4-body interactions. Therefore, $4$ or $6$ CNOT gates are necessary to implement a 3- or 4-body constraint, respectively. As the constraints are local interactions, there is no need to use SWAP gates which makes the parity embedding highly parallelizable \cite{Lechner2018}. In fact, for the circuit depth of arbitrary parity chip layouts, an upper bound independent of the system size is to be expected, while problems in the original all-to-all connected LHZ-architecture can always be implemented with constant depth \cite{Lechner2018}. This is of great importance when scaling up quantum devices to larger chips. 

\subsection{Parity QAOA with 4-body couplers}
When problems are plaquette compiled with the parity compiler \cite{compilerpaper}, it is also possible to implement each constraint as a single gate operation by using 4-body couplers as proposed by C. Dlaska et al.~\cite{rydbergpaper}. In this approach, every parity constraint is represented by a single 4-qubit gate with a fidelity comparable to that of a CNOT gate. With this strategy, the number of multi-qubit gates in the parity scheme will always be less than the number of CNOT gates needed in the standard gate model (without doing a parity transformation), even if no SWAP gates are taken into account. We will show this next. Let ${K>0}$ be the number of physical qubits (interactions) and ${N>0}$ the number of logical qubits in the problem. Let us further assume that we do not make use of ancilla qubits and that there are no symmetries in the logical graph and therefore Eq.~\eqref{eq:n_constraints} holds.
For the number $n_G$ of CNOT gates in the standard gate model, we assume that
\begin{equation}\label{eq:n_cnot_approx}
n_G \geq 2(K-N).
\end{equation}
This is justified by assuming that each interaction involving more than one qubit cannot be implemented with less than 2 CNOT gates. We subtract $N$, as there can occur (at most) $N$ single-qubit operations that do not involve CNOT gates in the standard gate model. It is not taken into account that some of the CNOT gates might be removed when optimizing the gate model circuit as this effect is assumed to be very small. The difference in CNOT gates is therefore
\begin{equation}
\Delta n = n_G-N_C \geq K-N = N_C
\end{equation}
which increases with the number of terms in the Hamiltonian (physical qubits). If the constraints can be implemented as single gate operations, the implementation in the standard gate model consumes (about) at least twice as many gates as the implementation using the parity compiler, according to this approximation. Note that the standard gate model still requires SWAP gates, which have not been considered in this estimate and significantly contribute to the number of CNOT gates. 

\section{Benchmarks}
In this section, we present benchmarks of the parity architecture against established methods to solve optimization problems on quantum computers. We benchmark the gate resources required for QAOA in the parity embedding as well as for directly embedding the problem Hamiltonian in the standard gate model with the same hardware requirements. For both architectures, a square lattice with connectivity between nearest neighbors is assumed in order to ensure a fair comparison. The number of CNOT gates (or 4-body couplers) is considered as the figure of merit. It is an indicator for the susceptibility to quantum noise, due to the fact that multi-qubit gates are more error-prone than single-qubit gates. The effects of real-world noise on the performance of QAOA are for example discussed in Refs.~\cite{Cheng2021, Marshall2020} or within the scope of the investigations in Ref.~\cite{Harrigan2021}, while the impact of quantum noise on QAOA in the parity architecture in particular is addressed in Refs.~\cite{rydbergpaper, hybridpaper}.

\subsection{Benchmark Preliminaries}
We deduce the number of required qubits and multi-qubit gates by making some assumptions rolled out next. In all our benchmarks, we assume the worst-case scenario for the parity mapping, namely that every constraint is a 4-body constraint and therefore consists of 6 CNOT gates. Thus, we provide a worst-case upper bound on the number $n_\text{CNOT}^\text{PM}$ of CNOT gates for a single QAOA-cycle in the parity mapping which is given by
\begin{equation}\label{eq:cnot_ratio}
    n_\text{CNOT}^\text{PM} = 6N_C.
\end{equation}
This bound is not reached in realistic circuits due to two main improvements that we do not consider here. On the one hand, the compiler considers not only 4-body constraints but also 3-body constraints (consuming only 4 CNOT gates). On the other hand, post processing on the circuit can allow to cancel some CNOT gates by rearranging the constraint circuits. The number of physical qubits is considered to be the number of terms in the logical Hamiltonian. This estimate neglects ancilla qubits in the parity scheme. Furthermore, we assume that there are no symmetries in the logical problem graph, i.e.\ Eq.~\eqref{eq:n_constraints} is valid.

For embedding QAOA in the standard gate model, we use the {t$\ket{\text{ket}}$} transpiler by Cambridge Quantum Computing (CQC) \cite{Sivarajah2020} to find a good mapping of qubits on a square-lattice chip and to place the SWAP gates automatically. This also takes into account the different ways to implement the phase gadgets occurring in the problem. In addition to applying SWAP gates, the {t$\ket{\text{ket}}$} transpiler also uses so-called bridge gates to perform CNOT operations between next-nearest neighbors on the square lattice more efficiently. A bridge gate consists of 4 CNOT gates.

Let us define the \textit{gate ratio} $r_\text{gates}$ 
\begin{equation}
    r_\text{gates}=\frac{n_\text{CNOT}(\text{PM})}{n_\text{CNOT}(\text{GM})}
\end{equation}
as the ratio between the required numbers of CNOT gates to implement one QAOA cycle in the parity scheme $n_\text{CNOT}(\text{PM})$ and in the standard gate model $n_\text{CNOT}(\text{GM})$.
If that ratio is below 1, the parity encoding is more efficient in terms of CNOT gates per QAOA cycle.

\subsection{Benchmarking $k$-body graphs}
We consider $k$-body graphs, each associated with a Hamiltonian $H$ containing only $\sigma_z$-terms, where each term is a product of at most $k$ Pauli $\sigma_z$-operators acting on different qubits. Note that according to this definition, a $k$-body graph is a hypergraph in the mathematical sense. The number of vertices corresponds to the number of (logical) qubits $N$ in the Hamiltonian $H$, while the number of (hyper-)edges is equivalent to the number of terms $K$ (which is the number of physical qubits in the parity scheme).

\begin{figure}
    \centering
    \includegraphics[width=.9\columnwidth]{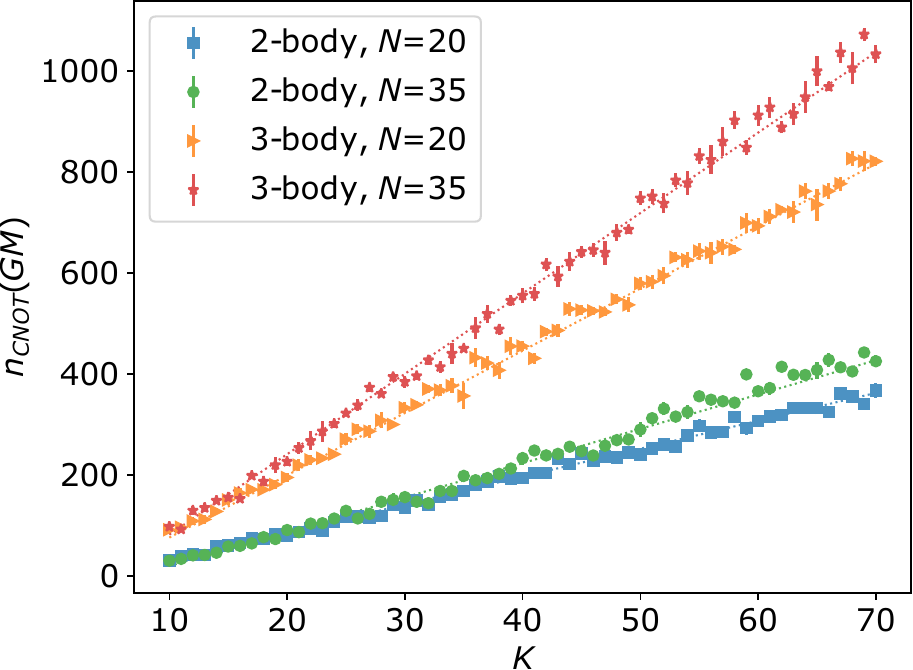}
    \caption{The number of CNOT gates required in the standard gate model in order to implement a QAOA cycle on a square lattice as a function of the number of interaction terms in the Hamiltonian. We show data for ${N=20}$ and ${N=35}$ to illustrate the different slopes of the linear fit functions. The error bars represent the standard deviation of the mean.}
    \label{fig:k-body:gates_vs_interactions}
\end{figure}

Let us consider a $k$-body graph with $N$ logical qubits. As in the previous section, let $n_G$ be the number of CNOT gates required to implement one QAOA cycle in the standard gate model on a nearest neighbor square lattice of size ${\sim \sqrt{N} \times \sqrt{N}}$. In the following, we investigate how $n_G$ depends on the number of $k$-body interactions in the problem Hamiltonian. To start, we fix $N$ and look at the special case of $k$-body graphs where only interactions of order $k$ (no lower orders) occur for ${k\in\{2, 3, 4, 5\}}$ and add a linear fit. For each system size $N$ we construct 10 instances with randomly chosen $k$-body couplings and take the average of the number of required CNOT gates. This number is determined by transpiling the circuit to a square lattice by using the {t$\ket{\text{ket}}$} transpiler. Exemplary data is visualized in Fig.~\ref{fig:k-body:gates_vs_interactions}. We observe that the number of required CNOT gates grows linearly with $K$. Furthermore, the slope of the linear fit grows with increasing $k$. This is in accordance with our expectations, as ${2k-1}$ CNOT gates are necessary to implement a $k$-body interaction (without considering SWAP gates) and higher-order interactions are more likely to be non-local, i.e.\ to require SWAP gates.

\begin{figure}
    \centering
    \includegraphics[width=.9\columnwidth]{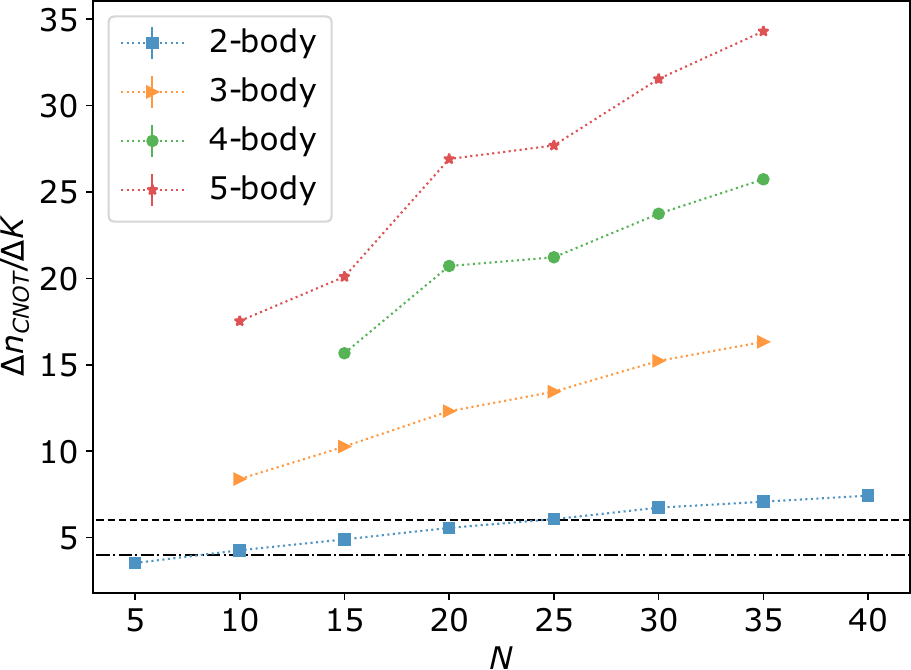}
    \caption{The average number of CNOT gates added per additional interaction term for different logical system sizes. Note that each data point corresponds to the slope of a linear fit such as in Fig.~\ref{fig:k-body:gates_vs_interactions}. The black dashed and dashed-dotted lines represent the worst (6 CNOT gates per constraint) and the best (4 CNOT gates per constraint) case that can occur in the parity mapping, respectively. Note that an additional interaction in the problem Hamiltonian usually introduces one additional constraint in the parity mapping if $N$ is constant.}
    \label{fig:k-body:gates_per_interaction}
\end{figure}

 We observe that the slope of the number of CNOT gates grows also with $N$, i.e.\ for larger systems it requires more gates to add additional interactions to the Hamiltonian. This is a consequence of the locality of the square lattice as for larger systems the number of interactions between non-neighboring qubits grows which in turn has to be compensated by SWAP gates and therefore additional CNOT gates. Fig.~\ref{fig:k-body:gates_per_interaction} depicts the slopes ${\Delta n_\text{CNOT}/\Delta K}$ of the linear fit functions to the CNOT curves with respect to $N$ for different interaction orders $k$. For each data point, logical systems with ${10\leq K \leq 70}$ $k$-body interactions are used. An exception is the point for ${N=5}$, where only ${\binom{5}{2}=10}$ 2-body interactions exist. Here, we used ${5\leq K \leq 10}$. For this small system, there are too few possibilities for higher-order interactions to obtain reliable data, which is why we only considered 2-body terms. For each physical system size, we analyze $5$ instances and take the average of the number of necessary gates. That is, we evaluate $305$ instances for each data point ($26$ instances for ${N=5}$). In the parity model, adding an interaction always introduces one additional constraint and therefore a constant number of CNOT gates. The results show that for small systems, it is more efficient to implement an additional 2-body interaction in the standard gate model than in the parity scheme. However, once higher-order interactions are included and for growing system sizes, it becomes advantageous to use the parity scheme. These results imply that the parity embedding shows a greater advantage 
\begin{enumerate}
    \item for larger logical system sizes and
    \item if more high-order terms (order ${k>2}$) are involved in the problem.
\end{enumerate}

These finding are also supported by the benchmarks on general $k$-body graphs, which contain a mix of different interaction types. We present these benchmarks next. 

\begin{figure}
    \centering
    \includegraphics[width=.9\columnwidth]{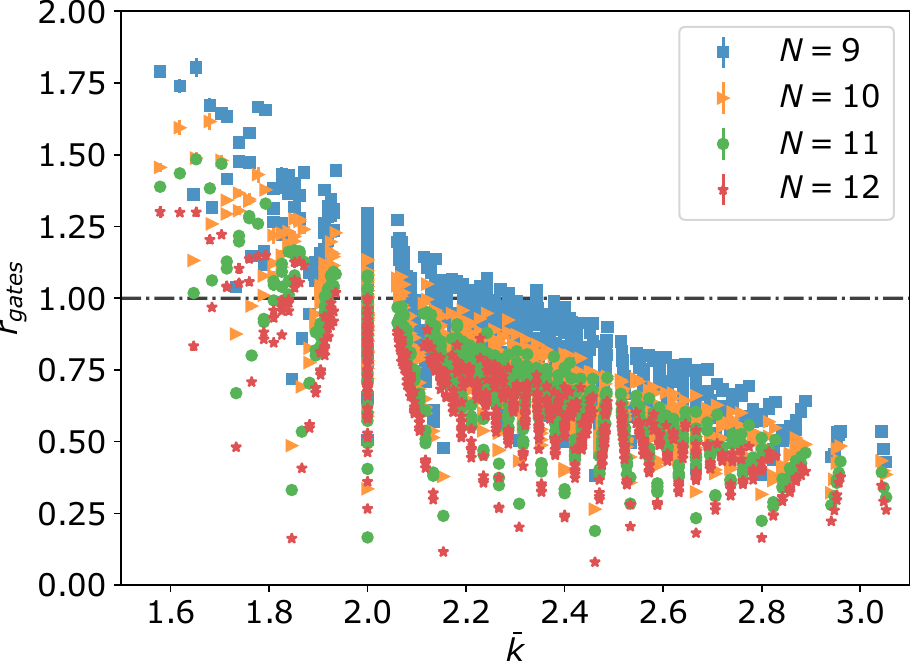}
    \caption{Benchmark results for different logical system sizes and graphs, constructed for different combinations of k-body interactions. We plot the gate ratio $r_\text{gates}$ against the mean interaction order $\bar{k}$. The dashed-dotted line indicates the gate ratio of 1, for data points below that, the parity architecture
    allows a more efficient embedding. The error bars denote the standard deviation of the mean and, in most cases, are smaller than the markers.}
    \label{fig:k-body:cnot_ratio_vs_interaction_order}
\end{figure}

We now compare the number of CNOT gates required to implement QAOA for a problem graph in the standard gate model with the number of gates when using the parity scheme (under the assumption that the constraints are implemented via CNOT gates). The number of logical qubits $N$ is chosen from ${N\in\{9, 10, 11, 12\}}$ and the number of terms $K$ is varied.
We also vary the number of occurrences of interactions of order $k$ (up to ${k=5}$), denoted by $n_k$.
The values for $n_k$ are chosen as follows: $n_{k=1}$ are taken from the set $\{0, 2, 4, 6, 8\}$, $n_{k=2}$ from the set $\{11, 13, 15, 17, 19\}$ and $n_{k=3,4,5}$ from the set $\{0, 2, 4\}$.
Note that ${K=\sum_{k=1}^{N} n_k}$. For all possible combinations of the given values, 10 random logical graphs were constructed and evaluated for the benchmarks in Fig.~\ref{fig:k-body:cnot_ratio_vs_interaction_order}, showing the gate ratio $r_\text{gates}$ versus the mean interaction order $\bar{k}$, given by
\begin{equation}\label{eq:mean_interaction_order}
\bar{k} := \frac{1}{K}\sum_{k=1}^{N} k n_k.
\end{equation} 

We observe that with a higher average interaction order the advantage of the parity architecture is more prominent. The data also implies that the advantage of the parity architecture increases with the number of logical qubits. However, this finding must be handled with care when only relying on that figure, due to the following reason. Increasing $N$ at constant $K$ removes constraints in the parity mapping and therefore CNOT gates. As we are using relatively sparse graphs, this effect may dominate when comparing different logical system sizes. Note that the benchmark presented in Fig.~\ref{fig:k-body:gates_per_interaction} does not exhibit this handicap, as we evaluate the number of gates that has to be added if an interaction (i.e.\ a physical qubit) is added while the logical system size is kept constant.

\subsection{The Financial Crash Problem}
A crucial problem in the financial industry is to predict how networks react to perturbations caused, for example, by a rapid drop of the market value of an institution \cite{Sornette2019, Estrella1998}. It is thus a natural question whether quantum computation can enhance forecasting financial crashes \cite{Orus2019}. This problem has already been investigated in terms of quantum annealing and also run on the D-Wave 2000 quantum annealer for a particular problem instance \cite{Ding2019}. Here, we use the same model for a QAOA benchmark. Let us first describe the model for completeness and then present benchmarks of the financial crash problem using the parity architecture.
\subsubsection{Financial Network Model}
A toy model for a financial network has been proposed in Ref.~\cite{Elliott2014}. In this model, a financial network is represented by $n$ financial institutions and $m$ assets. We define a vector $\mathbf{p}$ with $m$ entries, where the $k$-th entry $p_k$ represents the price of asset $k$. Further, the $n\times m$ matrix $\mathbf{D}$ defines the ownership, i.e.\ the element $D_{ij}$ corresponds to the percentage of asset $j$ owned by institution $i$. The matrix $\mathcal{C}$ is introduced as the holding matrix, i.e.\ the element $\mathcal{C}_{ij}$ is the percentage of institution $j$ held by institution $i$. For convenience, the matrix $\Tilde{\mathbf{C}}$ shall be defined as the diagonal matrix of self-holdings, such that ${\mathbf{C}= \mathcal{C}-\Tilde{\mathbf{C}}}$ only contains the cross-holdings. Following Ref.~\cite{ComputationalComplexityFinancialNetworks}, the equity values ${\mathbf{V}=(V_1, \dots, V_n)}$ of the institutions are defined as the solution of the linear system ${\mathbf{V}=\mathbf{D}\mathbf{p}+\mathbf{C}\mathbf{V}}$, such that ${\mathbf{V}=(\mathbb{1}-\mathbf{C})^{-1}\mathbf{D}\mathbf{p}}$. The market value $v_i$ of an institution $i$ is its equity value rescaled with its self-ownership. Therefore, the market values ${\mathbf{v}=(v_1, \dots, v_n)}$ can be written as 
 \begin{equation}
    \mathbf{v}=\Tilde{\mathbf{C}}(\mathbb{1}-\mathbf{C})^{-1}\mathbf{D}\mathbf{p}.
\end{equation}
In order to model a crash in the financial network, the functions ${b_i(v_i, \mathbf{p})=\beta_i(\mathbf{p})[1-\Theta(v_i-v_i^c)]}$ are introduced where $\Theta(x)$ denotes the Heaviside step function. Here, $v_i^c$ denotes a critical value for institution $i$. If the market value of institution $i$ drops below $v_i^c$, it causes a sudden drop in its equity value by $\beta_i(\mathbf{p})$ depending on the prices of assets. With that, the market values satisfy
 \begin{equation}\label{eq:fin_crash:equity_values}
    \mathbf{v}=\Tilde{\mathbf{C}}(\mathbb{1}-\mathbf{C})^{-1}[\mathbf{D}\mathbf{p}-\mathbf{b}(\mathbf{v}, \mathbf{p})],
\end{equation}
with ${\mathbf{b}=(b_1, \dots, b_n)}$. The term ${(\mathbb{1}-\mathbf{C})}$ is shown to be invertible in Ref.~\cite{ComputationalComplexityFinancialNetworks}. Due to the failure terms $b_i$, Eq.~\eqref{eq:fin_crash:equity_values} is highly non-linear. This makes the problem computationally hard to solve with classical algorithms. A financial network is said to be in equilibrium, iff it satisfies Eq.~\eqref{eq:fin_crash:equity_values}.
\subsubsection{Embedding in a quantum Hamiltonian}
We now seek for a quantum Hamiltonian that encodes the financial equilibrium in its ground state and follow the procedure given in Ref.~\cite{Orus2019}. The classical cost function
\begin{equation}\label{eq:fin_crash:cost}
    F(\mathbf{v}):=\left[\mathbf{v}-\Tilde{\mathbf{C}}(\mathbb{1}-\mathbf{C})^{-1}\left[\mathbf{D}\mathbf{p}-\mathbf{b}(\mathbf{v}, \mathbf{p})\right]\right]^2 \geq 0
\end{equation}
depending on the market values $\mathbf{v}$ has its global minimum at financial equilibrium and can be directly obtained by squaring Eq.~\eqref{eq:fin_crash:equity_values}.

The market values $v_i$ are now encoded in classical bits, which are then promoted to qubits, via the binary expansion ${v_i=\sum_{k=-\infty}^\infty x_{i,k}2^k}$ with ${x_{i,k}\in \{0, 1\}}$. The expression is truncated and restricted to integer numbers only, yielding 
\begin{equation}
    v_i \approx \sum_{k=0}^{q-1} x_{i,k}2^k,
\end{equation}
where $q$ is some positive integer number and represents the number of logical qubits per institution involved in the problem. This truncation results in an upper bound ${v^\text{max}=\sum_{k=0}^{q-1} 2^k}$ on the market values. 
In order to encode the Heaviside step function in a polynomial cost Hamiltonian, the function is rewritten in terms of the Fourier-Legendre expansion 
\begin{equation}\label{eq:fin_crash:heaviside_expansion}
    \Theta(x)=\frac{1}{2}+\sum_{l=1}^\infty\left[P_{l-1}(0)+P_{l+1}(0)\right]P_l(x), 
\end{equation}
which is valid in the interval $[-1, 1]$. Here, $P_l(x)$ denotes the $l$-th Legendre polynomial. We choose ${x=(v_i-v_i^c)/v^\text{max}}$ to ensure the argument of $\Theta(x)$ to be in the range where the expansion \eqref{eq:fin_crash:heaviside_expansion} is valid. If the expansion is truncated at level $r$, the failure functions ${b_i(v_i, \mathbf{p})}$ are only polynomials of degree $r$ in $v_i$. Therefore, the cost function $F(\mathbf{v})$ is a polynomial of degree $2r$ in $v_i$.

The last step is to map the bit variables $x_{i, k}$ to qubit operators $\hat{x}_{i,k}$ with eigenvalues ${\{0, 1\}}$, i.e.\ ${\hat{x}_{i,k}\ket{0}=0}$ and ${\hat{x}_{i,k}\ket{1}=\ket{1}}$. We write these operators in terms of the Pauli $\sigma_z$-operators and obtain
\begin{equation}
    \hat{x}_{i, k}=\frac{1+\sigma_z^{(i,k)}}{2}.
\end{equation}
With that mapping, the problem Hamiltonian is a polynomial of degree $2r$ in the Pauli $\sigma_z$ operators:
\begin{equation}
    H_p=\text{Poly}_{2r}(\sigma_z^{(i, k)}).
\end{equation}

\subsubsection{Problem instances}
We benchmark financial networks consisting of ${n\in \{3,4\}}$ institutions and ${m=7}$ assets. For promoting it to a quantum Hamiltonian according to the procedure discussed above, we use ${q=5}$ bits per institution and cut the Fourier-Legendre expansion at order ${r=3}$. In order to obtain reliable results, we perform benchmarks for $10$ instances of the financial network, where we randomly construct the ownership- and holding matrices. The minimum self-holding ratio of an institution was set to 0.5, and the assets' prices were chosen to range in the interval $[5, 20]$. The parameters $\beta_i$ were set to be 15\% of the equity values $V_i$ and the critical values $v_i^c$ were considered to be 80\% of the original market values $v_i$ (before perturbation). 

\subsubsection{Required resources}
The amount of qubit- and gate-resources required for a given number of institutions depends on the number of bits $q$ used to encode the market value of an institution as well as on the truncation parameter $r$. Both parameters limit the approximation accuracy of the model. Even for relatively resource-saving choices of these parameters, the number of terms is too large for an implementation on state-of-the-art devices. As an example, a problem with ${n=3}$, ${q=5}$, and ${r=3}$ already introduces ${n_\text{terms}^\text{tot}=1968}$ terms. Nevertheless, the absolute values of the coefficients vary by some magnitudes. That is, there are terms that dominate the problem while others may be neglected. This gives rise to the opportunity to save resources by only considering the terms with the dominating coefficients. In order to make the problem feasible for state-of-the-art devices, one might only keep the $n_\text{terms}^\text{trunc}$ terms with the most dominant coefficients and neglect others. Of course, this limits the accuracy of the solution and it is desirable to consider as many terms as possible. In that sense, $n_\text{terms}^\text{trunc}$ is also a parameter determining the accuracy of the calculation. Another possibility to omit terms with small coefficients is to introduce a threshold $c_\text{thres}$ (which we refer to as the \textit{chopping threshold}) and omit all terms with coefficients below this threshold.

\subsubsection{Benchmark Results}

\begin{figure}
    \centering
    \includegraphics[width=\columnwidth]{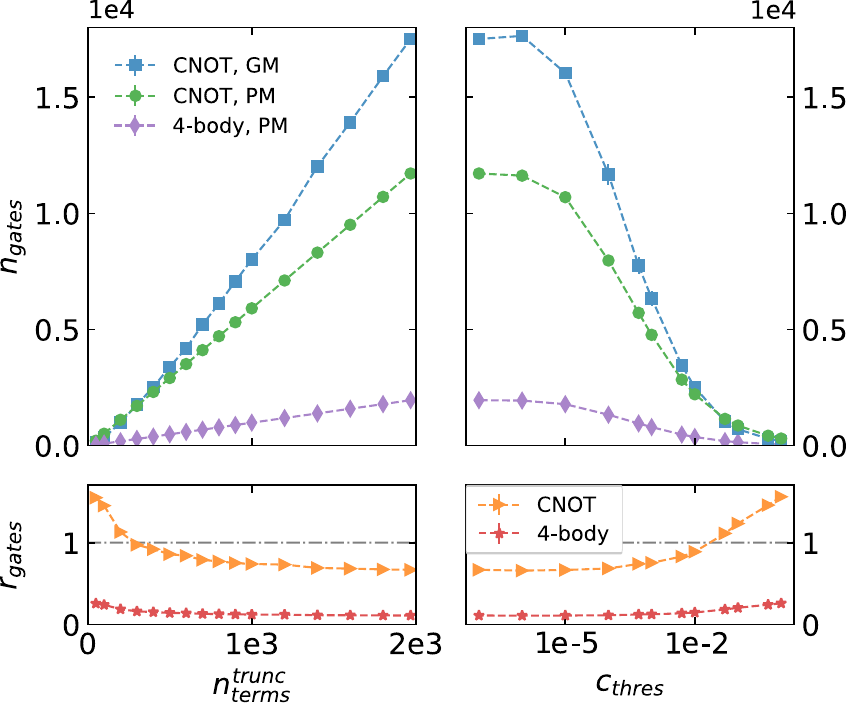}
    \caption{Benchmark results for 10 random financial networks with $3$ institutions, $7$ assets, and $5$ bits per institution. The Fourier-Legendre approximation was cut at order $3$. The average total numbers of CNOT gates and 4-body couplers (top panels) for the different embeddings are shown as well as their ratio (bottom panels). The dashed-dotted line indicates the gate-ratio of 1, for data points below that, the parity architecture allows a more efficient embedding. We vary the parameters $n_\text{terms}^\text{trunc}$ (left) and the chopping threshold $c_\text{thres}$ (right, note the logarithmic scale on the x-axis). In the extreme case, all terms of the Hamiltonian are considered. The error bars denote the standard deviation of the mean and are smaller than the markers in most cases.}
    \label{fig:fin_crash:benchmark_3inst}
\end{figure}

We compare the numbers of CNOT gates required for one QAOA-cycle in the parity embedding and in the standard gate model implementation with the methods described in the Benchmark Preliminaries. The results for ${n=3}$ institutions are depicted in Fig.~\ref{fig:fin_crash:benchmark_3inst}. Especially when taking many terms into account (which is equivalent to using a low chopping threshold), the parity embedding clearly outperforms the standard implementation when considering the number of CNOT gates as the figure of merit. We show the total number of CNOT gates as well as the ratio between CNOT gates in the parity mapping and in the standard gate model. Similar results are obtained by using financial networks with ${n=4}$ institutions (and leaving the other parameters invariant), which corresponds to increasing the number of logical qubits from 15 to 20. 

We also consider the case where the logical problem is parity compiled to a plaquette chip and the constraints are implemented by using 4-body couplers instead of CNOT gates. With that strategy, the advantage of the parity architecture is even more prominent. As we assume each constraint to require 6 CNOT gates in the standard parity embedding, this reduces the number of multi-qubit gates in the parity mapping by a factor of 6. As described above, the parity scheme can, with that approach, outperform the implementation in the standard gate model for all values of $n_\text{terms}^\text{trunc}$ and $c_\text{thres}$. This is also obvious from Fig.~\ref{fig:fin_crash:benchmark_3inst}.

\subsection{Electronic Structure Problems}
Finding the ground state of molecules is a fundamental challenge in contemporary molecular physics and quantum chemistry.  However, they are hard to tackle using state-of-the-art quantum computers, as they require a mapping of the problem to the spin model. This is problematic due to two main aspects. On the one hand, the electronic structure Hamiltonians (written in terms of spin operators) include not only $\sigma_z$-terms like the Ising model does.  On the other hand, they contain high-order terms, which have to be broken down to two-body interaction terms only in order to be implemented in the Ising model. R. Xia et. al. found a procedure to map the Hamiltonian to a Hamiltonian containing only (high-order) $\sigma_z$-terms by enlarging the Hilbert space \cite{Xia2018}. The minimum eigenvalue of this Hamiltonian corresponds to the ground-state energy of the initial Hamiltonian. The factor of $r$ introduced in Ref.~\cite{Xia2018} describes the number of replications of the Hilbert space and determines the quality of the approximation. We will outline this mapping in the next section. This idea of mapping the molecular Hamiltonian to an Ising problem has already been experimented with on the D-Wave quantum annealer \cite{Streif2018}. Seizing that approach, we use the paradigm of parity quantum computing to circumvent the translation to a Hamiltonian containing two-body interactions only. We benchmark the procedure of parity compiling the problem with the parity compiler \cite{compilerpaper} for performing QAOA against the direct implementation of the many-body interactions in the standard gate model, using only $\sigma_z$-terms.

\subsubsection{The Electronic Structure Hamiltonian}
The state of a multi-atomic molecule consisting of $M$ nuclei and $N$ electrons is determined by the mutual interaction of electrons and nuclei. The total Hamiltonian is given by 
\begin{equation}\label{eq:electronic_structure_hamiltonian}
    \begin{aligned}
         H=&-\sum_A \frac{1}{2M_A}\nabla^2_A - \sum_i \frac{1}{2}\nabla_i^2 + \sum_{A>B} \frac{Z_A Z_B}{|\mathbf{r}_A-\mathbf{r}_B|}- \\&-\sum_{A, i}\frac{Z_A}{|\mathbf{r}_A-\mathbf{r}_i|}+\sum_{i>j}\frac{1}{|\mathbf{r}_i-\mathbf{r}_j|},
    \end{aligned}
\end{equation}
where the upper-case indices refer to nuclei and lower-case indices to electrons. The vectors $\mathbf{r}$ represent the position vectors of the nuclei and electrons and $Z_A$, $Z_B$ and $M_A$ are the charge numbers and masses of the nuclei, respectively. By neglecting the kinetic energy term of the nuclei and their mutual interaction (Born-Oppenheimer approximation) and rewriting Eq.~\eqref{eq:electronic_structure_hamiltonian} in terms of the second quantization, we obtain
\begin{equation}
    H=\sum_{i, j} h_{ij}a_i^\dagger a_j +\frac{1}{2}\sum_{i, j, k, l}h_{ijkl}a_i^\dagger a_j^\dagger a_k a_l,
\end{equation}
where $a_i$ and $a_j^\dagger$ with ${\{a_i, a_j^\dagger\}=\delta_{ij}}$ are the fermionic annihilation and creation operators. The coefficients $h_{ij}$ and $h_{ijkl}$ are the one- and two-electron integrals for the basis set chosen. There are several procedures to map that form to a Hamiltonian containing Pauli operators. The most common ones are the Jordan-Wigner \cite{Jordan1928} and the Bravyi-Kitaev transform \cite{Bravyi2002}. After applying a transformation from the Hamiltonian in the second quantization formalism to a spin Hamiltonian, we arrive at a Hamiltonian of the form
\begin{equation}\label{eq:electronic_structure:spin_hamiltonian}
    \begin{aligned}
        H&=\sum_{i, \alpha} h_\alpha^i\sigma_\alpha^i + \sum_{ij \alpha\beta} h_{\alpha\beta}^{ij}\sigma_\alpha^i\sigma_\beta^j +\\ &+ \sum_{ijk \alpha\beta\gamma} h_{\alpha\beta\gamma}^{ijk} \sigma_\alpha^i\sigma_\beta^j\sigma_\gamma^k+ \dots,
    \end{aligned}
\end{equation}
where $\sigma_{\alpha}^i$ with ${\alpha\in \{x, y, z\}}$ represent the Pauli operators acting on qubit $i$. This Hamiltonian can now in principle be tackled on a gate-based quantum computer \cite{Seeley2012}. 

\subsubsection{Mapping the Electronic Structure Hamiltonian to $\sigma_z$-terms}
Such a mapping was proposed by R. Xia et al. in 2018 \cite{Xia2018}. They perform a mapping from a Hamiltonian containing arbitrary Pauli matrices to $\sigma_z$-interactions only, by replicating the $n$-qubit Hilbert space $r$ times, i.e.\ $rn$ ancilla qubits are introduced in the mapped logical space. A state ${\ket{\psi}=\sum_l \ket{\phi_l}}$ is mapped as
\begin{equation}
    \ket{\psi} \mapsto \ket{\Psi} = \bigotimes_l \bigotimes_{j=1}^{b_l} \ket{\phi_l},
\end{equation}
where $b_l$ is the number of times $\ket{\phi_l}$ is replicated. Note that ${r=\sum_l b_l}$.
The operators are mapped\footnote{We omitted the sign functions introduced in Ref.~\cite{Xia2018} as they are not relevant for our considerations.} with the relations 
\begin{align}
    \sigma_x^i &\mapsto \frac{1-\sigma_z^{i_j} \sigma_z^{i_k}}{2}\\
    \sigma_y^i &\mapsto \mathbf{i}\frac{\sigma_z^{i_k}-\sigma_z^{i_j}}{2} \\
    \sigma_z^i &\mapsto \frac{\sigma_z^{i_j}+\sigma_z^{i_k}}{2}\\
    \mathbb{1}^i &\mapsto \frac{1+\sigma_z^{i_j} \sigma_z^{i_k}}{2}
\end{align}
for acting on the $j$-th and $k$-th $n$-qubit subspace in $\ket{\Psi}$. 
The index $i_j$ refers to the $i$-th qubit in the $j$-th $n$-qubit space. 
To calculate expectation values one has to take all combinations of the $n$-qubit spaces $j$ and $k$ into account.
A more detailed discussion on these operators is conducted in Ref.~\cite{Xia2018}.
Note that this mapping approximates the original Hamiltonian and only captures its ground-state energy and not the full Hilbert space. With increasing $r$, more quantum effects are taken into account and the approximation error drops. We are now left with a Hamiltonian of the form
\begin{equation}\label{eq:electronic_structure:final_hamiltonian}
    H'=\sum_i f_i \sigma_z^i + \sum_{i,j} f_{ij} \sigma_z^i\sigma_z^j + \sum_{i, j, k} f_{ijk} \sigma_z^i\sigma_z^j\sigma_z^k+\dots,
\end{equation}
which has the same form as Hamiltonian \eqref{eq:logical_hamiltonian} and can be mapped to a QUBO problem or a parity encoding. The authors of Ref.~\cite{Xia2018} provide a procedure to find the ground-state energy of $H$ provided one is able to do so for $H'$.
We note that $H'$ can include an exponential number of terms, resulting in an exponential number of qubits in the parity mapping. In practice, the number of qubits can be reduced by truncating the Hamiltonian, as we did for the financial crash problem.

\subsubsection{Problem Instances}
For our benchmarks, we use the electronic structure Hamiltonians of $\text{H}_2$ and LiH. These molecules were also considered in a D-Wave experiment by M. Streif et al. in 2018 \cite{Streif2018}. For $\text{H}_2$, we consider both of the two molecular orbitals as active. However, when considering all six orbitals of LiH as active, we would be left with a 10-qubit Hamiltonian with hundreds of terms (even without mapping the spin Hamiltonian to $\sigma_z$-terms only). Due to limitations in computational resources, we therefore only consider three orbitals as active in the LiH molecule, which might not be enough to obtain reasonable results when really attempting to solve for the electronic structure of LiH, but still gives reasonable data for benchmarking the different architectures. The fermionic operators in the electronic structure Hamiltonians were mapped to qubit operators (Pauli operators) by using a binary code transform which also takes into account symmetries neglected by the Jordan-Wigner transform or the Bravyi-Kitaev transform, in order to reduce qubit requirements. For this procedure, the open-source Python library \texttt{openfermion} \cite{McClean2020} was used.

\subsubsection{Benchmark Results}
\begin{figure}
    \centering
    \includegraphics[width=\columnwidth]{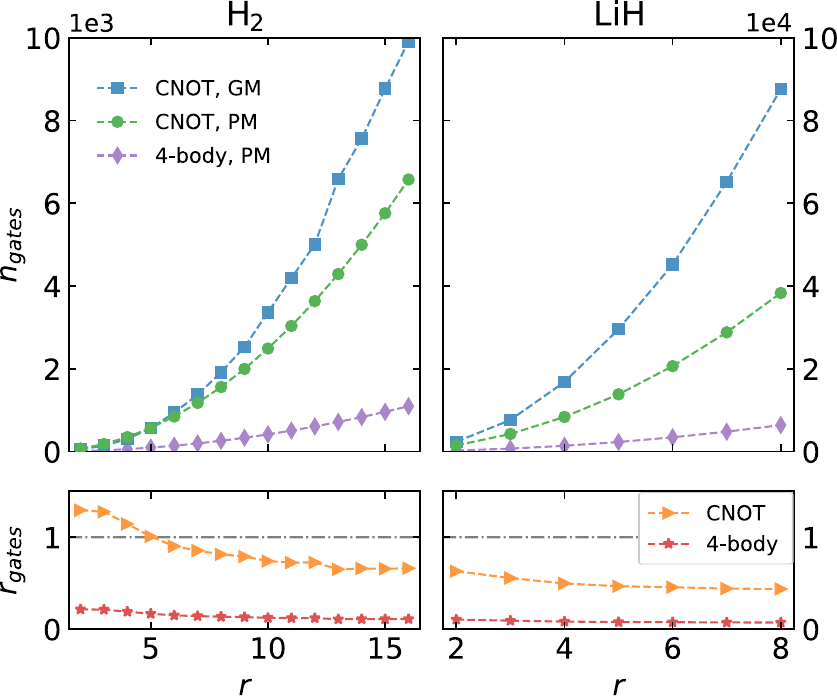}
    \caption{Benchmark results for the electronic structure Hamiltonians of $\text{H}_2$ (left) and LiH (right). We show results for the CNOT implementation of the parity compiled problem as well as for the implementation with native 4-body gates. Note that the standard gate model implementation always involves CNOT gates, as an implementation with 4-body gates is not possible. The absolute numbers of gates (top panels) are shown as well  as their ratio (bottom panels). The dashed-dotted line indicates the gate ratio of 1, for data points below that, the parity architecture allows a more efficient embedding.}
    \label{fig:electronic_stucture:all}
\end{figure}

We benchmark the parity mapping against the direct implementation of the problem Hamiltonian $H'$ in the standard gate model (obtained by transpiling the circuit with the {t$\ket{\text{ket}}$} transpiler). 
As we want to compare the required resources for QAOA in different protocols, we map it to a Hamiltonian with $\sigma_z$-operators only, in order to map it to an Ising spin glass (and subsequently embed it on a quantum annealer or perform QAOA) or to apply the parity transformation. Although applying approaches like the variational quantum eigensolver \cite{Peruzzo2014} directly to the Hamiltonian $H$ 
might require fewer resources as there is no need for replicating the Hilbert space, we use the mapping described above to synthetically benchmark QAOA in the parity architecture.
We compare the number of required multi-qubit gates for a single QAOA cycle. In Fig.~\ref{fig:electronic_stucture:all}, the results for the QAOA benchmark are shown for different values of the replication factor $r$, where the constraints in the parity mapping were implemented via CNOT gates as well as with 4-body couplers. The figures also show the ratio of required CNOT gates in the direct implementation vs. in the parity scheme as defined in Eq.~\eqref{eq:cnot_ratio} for different values of $r$. It is obvious that a threshold for the parity mapping to consume fewer CNOT gates exists at ${r=5}$ for $\text{H}_2$, while for LiH this threshold is already at ${r=2}$ (note that the trivial case ${r=1}$ is not shown in Fig.~\ref{fig:electronic_stucture:all}). For the implementation with constraint gates, the gate ratio is always significantly below 1.

The data presented here was obtained without performing any post-optimization of the circuit for the parity embedding. However, in this work no ancilla qubits (which might be necessary for compilation and would introduce additional constraints and therefore CNOT gates) were considered. With these assumptions, the parity architecture clearly outperforms the direct embedding of interactions in terms of CNOT gates, especially for large values of $r$.

Also for this problem type, the advantage of the parity embedding becomes even more prominent, if the parity constraints are not implemented in terms of CNOT gates, but by using 4-body couplers, as described above, which can also be seen from Fig.~\ref{fig:electronic_stucture:all}. Note that the number of 4-body couplers corresponds to the number of CNOT gates divided by 6, as we assume each constraint to consume 6 CNOT gates for the comparison of CNOT gates.
\setlength{\belowcaptionskip}{-10pt}

\section{Conclusion and Outlook}
We have presented benchmarks of the parity architecture for optimization problems that serve as toy models for highly relevant real-world scenarios. Our findings suggest a significant advantage of the parity embedding regarding gate resources on quantum devices with nearest-neighbor connectivity on a 2D-grid for the problems analyzed, especially for large system sizes. We demonstrate this advantage for devices with nearest neighbor CNOT interactions by comparing the parity architecture to state-of-the-art gate model compilers and for architectures that feature 4-body couplers. While there is already a strong reduction in number of gates when comparing CNOT gates, it improves further when implementing native 4-body couplers on the device. This could serve as a guideline for future hardware developments. 

The main hardware advantages of the parity mapping come on top of the pure embedding advantages which we analyzed here. This includes mainly the parallelizability of gates, i.e.\ the algorithm can be performed at constant circuit depth \cite{Lechner2018}, and with that the ability to design global gates where crosstalk can be mitigated by cancellation.
The results of our work in combination with further studies  (Refs.~\cite{rydbergpaper, hybridpaper}) show that the parity architecture is a promising candidate for developing scalable quantum optimization devices beyond the NISQ-era.\\

\textit{Acknowledgements -} We thank E. Solano and S. Mugel for discussing the financial crash model as well as R. Xia and S. Kais for discussing the replication method in the electronic structure problem. Work at the University of Innsbruck is supported by the European Union program Horizon 2020 under Grants Agreement No.~817482 (PASQuanS), and by the Austrian Science Fund (FWF) through a START grant under Project No. Y1067-N27 and the SFB BeyondC Project No. F7108-N38, the Hauser-Raspe foundation. This material is based upon work supported by the Defense Advanced Research Projects Agency (DARPA) under Contract No. HR001120C0068. Any opinions, findings and conclusions or recommendations expressed in this material are those of the author(s) and do not necessarily reflect the views of DARPA.
\bibliographystyle{quantum_abrv}

\onecolumngrid
\end{document}